\documentclass[aps,prl,twocolumn,showpacs]{revtex4}
\usepackage{graphicx}
\usepackage{epsfig}
\usepackage{dcolumn}
\usepackage{bm}
\begin{document}

\title{Non-equilibrium vertex correction:
disorder scattering in magnetic tunnel junctions}

\author{Youqi Ke$^1$, Ke Xia$^2$ and Hong Guo$^1$}

\affiliation{$^1$ Centre for the Physics of Materials and Department
of Physics, McGill University, Montreal, PQ, H3A 2T8, Canada\\$^2$
State Key Laboratory for Surface Physics, Institute of Physics,
Chinese Academy of Sciences, P.O. Box 603, Beijing 100080, China }
\date{\today }

\begin{abstract}
We report a first principles formalism and its numerical
implementation for treating quantum transport properties of
nanoelectronic devices with atomistic disorder. We developed a
nonequilibrium vertex correction (NVC) theory to handle the
configurational average of random disorder at the density matrix
level so that disorder effects to nonlinear and nonequilibrium
quantum transport can be calculated from atomic first principles in
a self-consistent and efficient manner. We implemented the NVC into
a Keldysh non-equilibrium Green's function (NEGF) based density
functional theory (DFT) and applied the NEGF-DFT-NVC formalism to
magnetic tunnel junctions with interface roughness disorder. Our
results show that disorder has dramatic effects to nonlinear spin
injection and tunnel magneto-resistance ratio.
\end{abstract}

\maketitle

Quantitative understanding of impurity effects is crucial for
nanoelectronics where device properties are strongly influenced by
or even built on such effects. Examples are electron scattering by
dopants in semiconductor nanowires\cite{SINW} and field effect
transistors, spin scattering by disorder in magnetic tunnel
junctions\cite{DMTJ}, and transport of spin polarised current in
dilute magnetic semiconductors\cite{ohno}. Unintentional impurities
sit inside a device at unpredictable locations and therefore, any
physical quantity predicted by theory should be averaged over
impurity configurations. In \emph{ab initio} calculations, one may
carry out this average by generating many impurity configurations
for a given concentration $x$, calculating the relevant physical
quantity for each configuration, and finally averaging the results.
Such a brute force calculation is often not practical for at least
two reasons. First, when $x$ is small as is typical in the case of
semiconductor devices, say 0.1\%, one would need a thousand host
atoms to accommodate just one impurity atom. Second, it is known
that impurity average may require huge number of
configurations\cite{Qiao}. These requirements make a calculation
prohibitively large. Considerable effort has therefore been devoted
in literature to develop approximate techniques which avoid brute
force. In this regard, a widely used technique is the coherent
potential approximation (CPA)\cite{P.Soven} as implemented in
KKR\cite{G.M.Stocks} and LMTO\cite{Turek_book} first principles
methods. So far CPA has been applied to \emph{equilibrium}
electronic structure and transport calculations\cite{carva2006}.
However, most nanoelectronic devices operate under
\emph{nonequilibrium} conditions, for instance one wishes to predict
nonlinear current-voltage (I-V) characteristics. It is thus very
important to develop appropriate nonequilibrium techniques for
impurity averaging.

Here we report our solution of the atomistic \emph{nonequilibrium}
impurity average problem for quantum transport. We start from a
state-of-the-art real space atomistic quantum transport formalism
where density functional theory (DFT) is carried out within the
Keldysh nonequilibrium Green's function (NEGF)
framework\cite{mcdcal,Datta-book}. The basic idea of NEGF-DFT is
that the device Hamiltonian and electronic structure are determined
by DFT, the non-equilibrium quantum statistics of the device physics
is determined by NEGF, and the transport boundary conditions under
external bias are handled by real space numerical technique. We deal
with impurity average at single particle retarded Green's function
level by CPA\cite{P.Soven}, and at NEGF level by evaluating a
nonequilibrium vertex correction (NVC) term. The NEGF-DFT-NVC
formalism allows us to construct \emph{nonequilibrium density
matrix} self-consistently that includes impurity averaging. We then
apply our NEGF-DFT-NVC formalism to investigate effects of interface
roughness disorder in a magnetic tunnel junction (MTJ). Our results
indicate that disorder effect can drastically and qualitatively
influence nonlinear I-V curves and tunnel magneto-resistance ratio.

\begin{figure}[tbh]
\includegraphics[width=5cm]{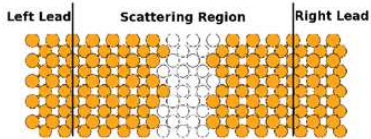}
\caption{(Colour online) Schematic of atomic structure of the
Fe/Vac/Fe magnetic tunnel junction. The two Fe/Vac interfaces have
roughness disorder. Fe: yellow spheres; vacuum: white spheres.}
\label{fig1}
\end{figure}

We consider a two-probe device consisting of a scattering region and
two semi-infinite leads extending along the transport direction {\bf
z} to $z=\pm \infty$, as shown in Fig.\ref{fig1}. The
system is periodically extended along the transverse (x,y)
direction. Note the scattering region includes several layers of
lead atoms\cite{mcdcal}. A bias voltage $V_b$ is applied across the
leads to drive a current flow, \emph{i.e.} $\mu_l-\mu_r = eV_b$
where $\mu_{l,r}$ are electrochemical potentials of the left/right
leads. We assume that impurities exist inside the scattering region
randomly but not in the leads. We further assume that any atomic
position $R$ in the scattering region may be occupied by two atomic
species, the host and impurity atoms labelled by $Q=A,B$ with
concentrations $C_R^A$ and $C_R^B$ such that
$C_{R}^{A}+C_{R}^{B}=1$.

In a NEGF-DFT self-consistent analysis\cite{mcdcal} of
\emph{ordered} systems, the non-equilibrium density matrix is
calculated by NEGF ${\bf G}^<(E)$, \emph{i.e.} $\hat{n}(E) \sim {\bf
G}^<(E)$, here $G^<$ satisfies the Keldysh equation $G^< =
G^{\cal{R}}\Sigma^< G^{\cal{A}}$ where
$G^{\cal{R}}=(G^{\cal{A}})^\dagger$ is the retarded Green's
function. $\Sigma^<= i\Gamma_{l}f_l + i\Gamma_{r}f_r$ is the lesser
self-energy where $f_{l,r}$ are Fermi functions of the left/right
leads; $\Gamma_{l,r}$ are line-width functions describing coupling
of the scattering region to the leads and they can be calculated by
standard iterative methods\cite{mcdcal}. Translational invariance of
ordered system allows one to evaluate all quantities in the unit
cell by integration over two dimensional (2D) Brillouin zone (BZ) in
the (x,y) direction. When there are impurities, translational
symmetry is broken. The spirit of CPA is to construct an effective
medium theory by an impurity configurational average that restores
the translational invariance. For NEGF this means calculating
$\overline{G}^< = \overline{G^{\cal{R}} \Sigma^< G^{\cal{A}}}$. Even
though there is no impurity in the leads to affect $\Sigma^<$, the
impurity average $\overline{(\cdots)}$ correlates Green's functions
$G^{\cal{R}}$ with $G^{\cal{A}}$. In particular, $\overline{G}^< \ne
\overline{G}^{\cal{R}} \Sigma^< \overline{G}^{\cal{A}}$ due to
multiple scattering by the impurities. To calculate
$\overline{G}^<$, we introduce a self-energy $\Gamma_{NVC}$ that is
a consequence of impurity scattering at \emph{nonequilibrium}, such
that $\overline{G}^< =\overline{G}^{\cal{R}} (\Sigma^<+\Gamma_{NVC})
\overline{G}^{\cal{A}}$. $\Gamma_{NVC}$ is called
\emph{nonequilibrium vertex correction} (NVC) whose equilibrium
counterpart is well known in calculations of Kubo formula by Feynman
diagrammatic techniques\cite{Mahan}. There is however a major
qualitative difference here: $\Gamma_{NVC}$ depends on the
non-equilibrium quantum statistical information of the device
scattering region while the equilibrium VC does not.

We found that $\Gamma_{NVC}$ is most conveniently calculable using a
\emph{site} oriented calculation scheme, for this reason we develop
our NEGF-NVC theory within TB-LMTO DFT implementation\cite{TB-LMTO},
using CPA \cite{Turek_book,carva2006}to describe the averaged
system. In this approach, impurity average for any single site
physical quantity $X_R$ is given by
\begin{equation}
\overline{X}_R\ =\ \sum_{Q=A,B} C_R^Q \overline{X}_R^Q
\label{Ave1}
\end{equation}
where $\overline{X}_R^Q$ is the conditional average over a
particular atomic specie $Q$ at site $R$ which is calculated by
\begin{equation}
\overline{X}_R^Q \ =\ \overline{\eta_R^Q X_R}/C_R^Q \ . \label{Ave2}
\end{equation}
Here $\eta_R^Q$ is the occupation of site $R$ by atomic specie $Q$
and its average is $\overline{\eta}_R^Q=C_R^Q$. Eq.(\ref{Ave1})
means the average of a quantity at site $R$ is a linear combination
of contributions from each atomic species.

The technical derivation details are given in the Supplemental
Material associated with this paper\cite{EPAPS}, here we briefly
outline the spirit of the theory. The impurity average of the site
diagonal NEGF is carried out by application of Eq.(\ref{Ave1}), {\it
i.e.} $\overline{G}_{RR}^<=\sum\limits_{Q=A,B}C_R^Q
\overline{G}_{RR}^{<,Q}$. The \emph{atom resolved} NEGF
$\overline{G}_{RR}^{<,Q}$ gives atom resolved average local charge
density $\overline{n}_R^Q \sim \overline{G}_{RR}^{<,Q}$ which is
needed in the DFT self-consistent iterations\cite{mcdcal}. To find
$\overline{G}_{RR}^{<,Q}$, we use Eq.(\ref{Ave2}). The final
expressions of $\overline{G}_{RR}^{<,Q}$ are given in Eqs.(10,28,29)
of the Supplemental Material\cite{EPAPS}, and it is related to
$\Gamma_{NVC}$ of the auxiliary NEGF. The calculation of NVC is
carried out by application of single-site approximation (SSA) within
CPA-based multiple scattering theory as summarised in Supplemental
Material\cite{EPAPS} where the final expression is given by Eq.(23)
there. From $\Gamma_{NVC}$ we obtain $\overline{G}_{RR}^{<,Q}$ hence
the averaged density matrix $\overline{n}_{R}^Q$ for atom $Q$ on
site $R$. The charge density is used to calculate device Hamiltonian
for the next step in the DFT iteration and this procedure is
repeated until numerical convergence.

An extremely stringent test of our NEGF-DFT-NVC formalism and its
numerical implementation is carried out by calculating two-probe
devices at \emph{equilibrium}, and check if the
fluctuation-dissipation relationship is satisfied or not.
Mathematically, fluctuation-dissipation theorem dictates
$\overline{G}_{RR}^{<,Q}=\overline{G}_{RR}^{{\cal{A}},Q}-
\overline{G}_{RR}^{{\cal{R}},Q}$ at \emph{equilibrium}. Here,
calculation of $\overline{G}_{RR}^{<,Q}$ requires NVC while
calculations of $\overline{G}_{RR}^{{\cal{A}},Q}$ and
$\overline{G}_{RR}^{{\cal{R}},Q}$ do not. For many disordered device
structures including that in Fig.\ref{fig1} we have checked, the
fluctuation-dissipation relationship is always satisfied to at least
one part in a million and the final tiny difference can be
attributed to numerical calculation issues. Importantly, we found
that NVC is extremely important: without it the density matrix and
transmission coefficients can have large errors and even become
qualitatively incorrect.

After the NEGF-DFT-NVC self-consistent calculation is converged we
calculate current-voltage (I-V) characteristics by Landauer formula,
where an additional vertex correction must be done on the
transmission coefficients\cite{Kudrnovsky}. At low temperature the
I-V curve is given by:
\begin{equation}
\label{eq-6} \overline{I}=\frac{e}{h}\int\nolimits_{\mu _{r}}^{\mu
_{l}}Tr\left[\overline{\Gamma_l g^{\cal{R}}\Gamma
_rg^{\cal{A}}}\right]dE\ .
\end{equation}
The integrand of the above expression is the transmission
coefficient where the impurity average, once again, correlates the
retarded $g^{\cal{R}}$ and advanced $g^{\cal{A}}$ Green's functions
($g^{\cal{R,A}}$ are auxiliary Green's functions of $G^{\cal{R,A}}$,
see Supplemental Material) connected by line-widths $\Gamma_{l,r} $
corresponding to the auxiliary $ g $. We write the averaged
transmission into a coherent part and a vertex part which describes
the inter-channel scattering events:
\begin{eqnarray}
\label{eq-7} \overline{T} &=&Tr\left[ \Gamma_l\overline{g}^{\cal{R}}
\Gamma_r\overline{g}^{\cal{A}}\right] + Tr\left[ \Gamma_l
\overline{g}^{\cal{R}} \Gamma_{VC}^{\prime} \overline{g}^{\cal{A}}
\right]
\end{eqnarray}
where $\Gamma _{VC,R}^{\prime }$ is obtained from the expression of
$\Gamma_{NVC}$ by replacing $\Sigma^{<}$ with $\Gamma_{r}$. The
equilibrium conductance for a spin channel is given by
$\overline{T}(e^2/h)$.

As an important application of the NEGF-DFT-NVC formalism, we have
investigated a disordered MTJ shown in Fig.\ref{fig1} which consists
a vacuum (Vac) tunnel barrier sandwiched by two Fe leads. The
$Fe/Vac$ interface has roughness disorder which substantially
influences spin dependent transport at equilibrium\cite{Xu}. Here we
focus on nonequilibrium. In our Fe/Vac/Fe MTJ, the left/right Fe/Vac
interface layers has $x$\% and $(1-x)$\% Fe atoms respectively, and
the rest are vacuum sites. The scattering region consists of ten
perfect atomic layers of Fe oriented along (100) on the left and
right ending with the rough interface sandwiching four vacuum
layers. The scattering region is connected to perfect Fe left/right
leads extending to $z=\pm \infty$. Because the entire structure,
after impurity average, is periodic along the transverse $x,y$
directions, we found that very careful two-dimensional Brillouin
zone (BZ) sampling is necessary in calculating the density matrix:
we use $200\times 200$ k-mesh to ensure excellent numerical
convergence. For the I-V curve calculation, Eq.(\ref{eq-6}),
$300\times 300$ BZ k-mesh is used for each point in the energy
integration. For other DFT details we follow standard TB-LMTO
literature\cite{Xu}.

\begin{figure}[tbh]
\includegraphics[width=8.2cm]{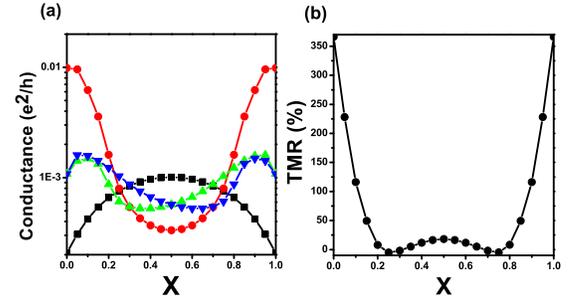}
\caption{(Colour online) (a) Conductance $G_{\uparrow,\downarrow}$
versus disorder $x$ at equilibrium. Red circles: $G_\downarrow$ in
PC; black squares: $G_\uparrow$ in PC. Blue down-triangles:
$G_\downarrow$ in APC; Green up-triangles: $G_\uparrow$ in APC. (b)
TMR Versus $x$. } \label{fig2}
\end{figure}

Fig.\ref{fig2}a is a semi-log plot of equilibrium ($V_b=0$)
conductance versus disorder $x$ for both spin-up and -down channels,
$G_{\uparrow},G_{\downarrow}$. The four curves correspond to
magnetic moments of the two Fe leads having parallel or
anti-parallel configurations (PC or APC). Since the left interface
is chosen to be $Fe_xVac_{1-x}$ while the right $Fe_{1-x}Vac_x$,
$G_{\uparrow, \downarrow}$ are both symmetric about $x=0.5$ in PC
(black squares and red circles). For APC they are not symmetric but
satisfy $G_{\uparrow}(x)=G_{\downarrow}(1-x)$, as expected. Impurity
scattering dramatically decreases $G_\downarrow$ in PC, and has
relatively less effect for $G_{\uparrow}$ and for APC. It was well
known\cite{surface-state} that for perfect interfaces, the surface
electronic states of Fe give resonance transmission. These
resonances are destroyed rapidly by the interface disorder as $x$
changes from zero to 50\% leading to the drastic reduction of
$G_{\downarrow}$ in PC. An important device merit for MTJ is the
tunnel magneto-resistance ratio (TMR) defined by total tunnelling
currents for PC and APC: TMR=$(I^{PC} -I^{APC})/I^{APC}$. At
equilibrium when all currents vanish, we use equilibrium
conductances to calculate TMR. Fig.\ref{fig2}b plots equilibrium TMR
versus $x$ showing a dramatic effect of disorder. In particular, TMR
drops to very small values, even to slightly negative values, as $x$
is increased from zero. These equilibrium features are consistent
with previous super-cell calculations\cite{Xu}.

\begin{figure}[tbh]
\includegraphics[width=8.2cm]{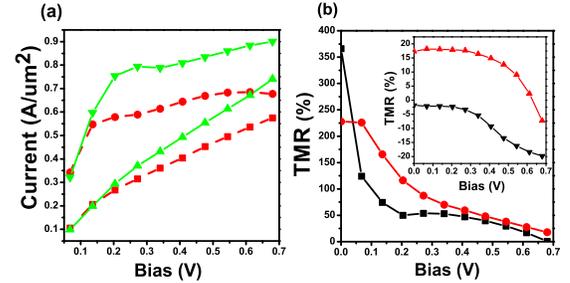}
\caption{(Colour online) (a) Comparison of I-V curves with disorder
$x=0.05$. Solid lines (green): current for PC (up-triangles) and APC
(down-triangles) without using NVC in density matrix self-consistent
iteration. Dashed lines (red): current for PC (circles) and APC
(squares) using the full NVC formalism. (b) TMR versus bias voltage
$V_b$ for four different values of $x$. The main figure is for x=0.0
(black squares) and x=0.05 (red circles); the inset for x=0.3 (black
down-triangles) and x=0.5 (red up-triangles) } \label{fig3}
\end{figure}

We now investigate nonequilibrium properties when $V_b\neq 0$  so
that current flows through. To show the importance of NVC, we have
calculated I-V curves at $x=0.05$ by including vertex correction
only at the level of transmission coefficient, {\it i.e.} without
NVC in the NEGF-DFT self-consistent iterations of the density
matrix: the solid lines (green) in Fig.\ref{fig3}a plot this result.
In comparison, the dashed lines (red) plot the full results where
NVC is included. The substantial differences indicate that NVC is
extremely important for obtaining correct results at nonequilibrium.
Fig.\ref{fig3}b and its inset plot TMR versus bias for four values
of $x=0.0, 0.05, 0.3, 0.5$, obtained by the full NVC formalism. For
zero or small values of $x$, TMR reduces with $V_b$ as is often seen
in experimental measurements\cite{yuasa}. For larger $x$, for
instance $x \sim 0.5$, TMR can go negative as $V_b$ is increased.
Indeed, experimental measurements had seen\cite{moodera} negative
TMR at large $V_b$, although for different MTJs and possibly
different physical origin. Very dramatically, at $x=0.3$, the entire
TMR curve is negative: here the absolute value of TMR actually
increases with $V_b$ (see inset). These behaviours of TMR strongly
suggest that interface disorder play very important roles for
nonequilibrium spin injection.

\begin{figure}[tbh]
\includegraphics[width=8.2cm]{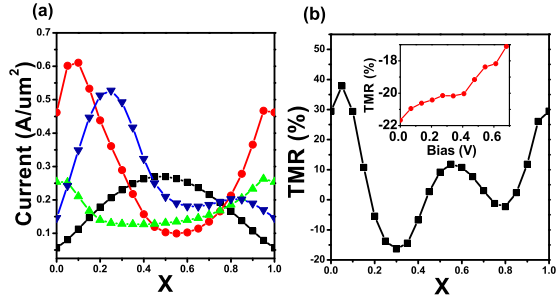}
\caption{(Colour online) (a) Spin currents versus disorder $x$ at
bias $V_b=0.544$V, for PC and APC. Red circles and black squares:
spin currents for spin-up and -down in PC; green up-triangles and
blue down-triangles: spin currents for spin-up and -down in APC. (b)
TMR versus $x$ at the same $V_b$. Inset of (b): TMR versus $V_b$ for
a device where left and right interfaces have different values of
$x$, on the left interface $x=0.3$, on the right $x=0.05$. }
\label{fig4}
\end{figure}

Fig.\ref{fig4} plots spin currents and TMR versus disorder $x$ at
$V_b=0.544$V. This is to be compared with Fig.\ref{fig2} where
$V_b=0$. A finite bias breaks left-right symmetry of the atomic
structure and therefore, the spin currents do not have a symmetric
behaviour about $x=0.5$ anymore. Both spin currents (Fig.\ref{fig4}a)
and TMR (Fig.\ref{fig4}b) varies with disorder $x$ in substantial
ways. In particular, TMR rapidly dips to negative values when $x$ is
increased to about 20\%. So far we have focused on devices where the
left has a Fe$_x$Vac$_{1-x}$ interface while the right has
Fe$_{1-x}$Vac$_x$. We have also applied the NEGF-DFT-NVC formalism
to devices where the left and right interfaces are disordered
totally differently. The inset of Fig.\ref{fig4}b plots TMR for such
a system where left interface has $x=0.3$ while the right interface
has $x=0.05$. For this system TMR is negative and its absolute value
decreases as $V_b$ is increased which is qualitatively similar to
what discussed above.

In summary, we have developed a nonequilibrium vertex correction theory
and its associated software for analysing quantum transport properties of
disordered nanoelectronic devices at nonequilibrium. The impurity averaging
of the nonequilibrium density matrix is facilitated by the NVC self-energy
that is related to quantum statistical information of the device scattering
region. Our NEGF-DFT-NVC theory has several desired features, including
atomistic first principle, non-equilibrium, efficient configurational average
and self-consistency. This allows us to analyse nonequilibrium quantum
transport of realistic device structures including realistic atomic
substitutional impurities. Using this tool, we have calculated nonlinear
spin currents in Fe/Vac/Fe MTJ with interface roughness disorder, and found
that effects of NVC can play a dominant role in determining the properties
of spin injection.

{\bf Acknowledgements.} We gratefully acknowledge useful discussions
with Zhanyu Ning, Wei Ji, Dr. Lei Liu and Dr. Eric Zhu. This work is
supported by NSERC of Canada, FQRNT of Quebec and CIFAR (H.G.). K.X.
is supported by NSF-China (10634070) and MOST (2006CB933000,
2006AA03Z402) of China. We are grateful to RQCHP for providing computation
facility.

\newpage

\end{document}